\documentclass[aps,prd,onecolumn,groupedaddress,showpacs,nofootinbib,amssymb]{revtex4}
\usepackage[dvips]{graphicx}
\usepackage{amssymb}
\usepackage{amsmath}
\usepackage{graphicx}
\usepackage{amsfonts}
\usepackage{bm}

\newcommand{\be}{\begin{equation}}
\newcommand{\ee}{\end{equation}}
\newcommand{\bea}{\begin{eqnarray}}
\newcommand{\eea}{\end{eqnarray}}
\newcommand{\beaa}{\begin{eqnarray*}}
\newcommand{\eeaa}{\end{eqnarray*}}

\newcommand{\e}{\mathrm{e}}

\begin{document}

\title{Unimodular Mimetic $F(R)$ Inflation}

\author{S.~D.~Odintsov$^{1,2,3}$}
\email{odintsov@ieec.uab.es}
\affiliation{$^{1)}$Institut de Ciencies de lEspai (IEEC-CSIC),
Campus UAB, Carrer de Can Magrans, s/n\\
08193 Cerdanyola del Valles, Barcelona, Spain}
\affiliation{$^{2)}$ICREA, Passeig LluA­s Companys, 23,
08010 Barcelona, Spain}
\affiliation{ $^{3)}$Tomsk State Pedagogical University, 634061 Tomsk, Russia}
\author{V.~K.~Oikonomou$^{4,5}$}
\email{v.k.oikonomou1979@gmail.com, voiko@sch.gr}
\affiliation{ $^{4)}$Laboratory for Theoretical Cosmology, Tomsk State University of Control Systems
and Radioelectronics (TUSUR), 634050 Tomsk, Russia}
\affiliation{ $^{5)}$National Research Tomsk State University, 634050 Tomsk,
Russia}

\begin{abstract}
We propose the unimodular-mimetic $F(R)$ gravity theory, to resolve cosmological constant problem and dark matter problem in a unified geometric manner. We demonstrate that such a theory naturally admits accelerating universe evolution. Furthermore, we construct unimodular-mimetic $F(R)$ inflationary cosmological scenarios compatible with the Planck and BICEP2/Keck-Array observational data. We also address the graceful exit issue, which is guaranteed by the existence of unstable de Sitter vacua.
\end{abstract}


\maketitle

\section{Introduction}

The cosmological constant problem \cite{Peebles:2002gy} along with the dark matter issue, are two of the unanswered problems in modern theoretical physics. With regards to the dark matter issue, it is believed that dark matter is possibly described by a particle which is weakly interacting with ordinary matter, see \cite{oikonomoudark} for an account on the particle nature of dark matter. However, for both these two problems, there exist two theoretical proposals in the context of generalized general relativity, namely the unimodular gravity \cite{unimodular,unimime,unibounce,nojiriuni} and mimetic gravity \cite{mukhanov1,mukhanov2,Golovnev,NO2,mimetic1,mimeletter,mimeticblackhole,mimeticbig}.  The unimodular gravity was firstly introduced to solve the cosmological constant problem in a consistent and elegant way, since the cosmological constant arises from the trace-free part of the Einstein field equations. Particularly, the trace-free part in unimodular gravity is obtained by imposing the following constraint on the metric determinant $\sqrt{-g}=1$. The mimetic gravity approach exploits the internal conformal degrees of freedom of the metric, and in result, a dark matter description is obtained, without the need for a perfect fluid to be present.

In a recent work \cite{unimime}, we combined the unimodular gravity and mimetic gravity disciplines to a unified framework in the context of Einstein-Hilbert gravity, and we were able to realize various cosmological scenarios which were exotic for the ordinary Einstein-Hilbert gravity, without the need  of extra matter fluids. In this paper we shall extend the formalism of unimodular-mimetic gravity of Ref. \cite{unimime}, in the context of $F(R)$ gravity (for reviews see \cite{reviews1}) and we study how various cosmological scenarios can be described by vacuum unimodular-mimetic $F(R)$, to which we shall refer to as U-M $F(R)$ gravity for brevity. The unimodular and mimetic gravity constraints shall be realized in the case at hand, by using the Lagrange multiplier method \cite{CMO}. Note that such theory is expected to solve the dark matter problem and the dynamical cosmological constant problem in the unified geometric formulation.

Particularly, we shall investigate how de Sitter cosmology and how the cosmology corresponding to a perfect fluid with constant equation of state (EoS) parameter $w$ are realized. In addition, by using the prefect fluid approach \cite{perfectfluid}, we shall demonstrate that it is possible to realize inflationary cosmologies \cite{inflation} with U-M $F(R)$ gravity, which are compatible to both the latest Planck \cite{planck} and BICEP2/Keck-Array data \cite{bicep2}. Moreover, we address the graceful exit issue in the context of U-M $F(R)$ gravity, and as we demonstrate, the graceful exit from inflation is guaranteed by the existence of unstable de Sitter vacua.

This paper is organized as follows: In section II, we present the formalism of U-M $F(R)$ gravity by using the Lagrange multiplier method, and we briefly investigate how various cosmological scenarios can be realized in the context of U-M gravity. In section III, we use the perfect fluid approach to realize with U-M $F(R)$ gravity an inflationary cosmology compatible with the observational data and in section IV we demonstrate how graceful exit can be achieved in the context of U-M $F(R)$ gravity. The conclusions follow in the end of the paper.

\section{Unimodular Mimetic $F(R)$ Gravity Using the Lagrange-multiplier Formalism}

In the Einstein-Hilbert mimetic gravity, the physical metric $g_{\mu \nu}$ is written in terms of an auxiliary scalar field and in terms of an auxiliary metric $\hat{g}_{\mu \nu}$, as follows,
\begin{equation}\label{metrpar}
g_{\mu \nu}=-\hat{g}^{\mu \nu}\partial_{\rho}\phi \partial_{\sigma}\phi
\hat{g}_{\mu \nu}\, .
\end{equation}
By using the representation of Eq. (\ref{metrpar}), results in the following constraint equation,
\begin{equation}\label{mimeticconstraint}
g^{\mu \nu}(\hat{g}_{\mu \nu},\phi)\partial_{\mu}\phi\partial_{\nu}\phi=-1\,
,
\end{equation}
to which we shall refer to as the mimetic constraint. In addition, in Einstein-Hilbert unimodular gravity, the following constraint holds true,
\be
\label{unimodularconstraint}
\sqrt{-g}=1\, ,
\ee
to which we shall refer to as the unimodular constraint. Before we continue, let us briefly describe the background geometry, which we shall assume to be described by a flat Friedman-Robertson-Walker (FRW) metric, with line element,
\begin{equation}\label{frw}
ds^2 = - dt^2 + a(t)^2 \sum_{i=1,2,3}
\left(dx^i\right)^2\, ,
\end{equation}
with $a(t)$ being the scale factor. In addition, the Ricci scalar corresponding to the flat FRW metric of Eq. (\ref{frw}), is equal to,
\begin{equation}\label{ricciscalarnnew}
R=6\left (\dot{H}+2H^2 \right )\, .
\end{equation}
For simplicity we assume that the scalar field $\phi$ is only time-dependent, so $\phi=\phi (t)$. In order to realize the mimetic and unimodular constraints of Eqs. (\ref{mimeticconstraint}) and (\ref{unimodularconstraint}), we shall make use of the Lagrange multiplier formalism of Refs. \cite{CMO}, and we introduce two Lagrange multipliers in the $F(R)$ gravity action, which we denote $\eta$ and $\lambda$. The Lagrange multiplier $\eta$ corresponds to the mimetic constraint (\ref{mimeticconstraint}) and the Lagrange multiplier $\lambda$ corresponds to the unimodular constraint (\ref{unimodularconstraint}), so the $F(R)$ gravity action with potential and Lagrange multipliers is equal to,
\begin{equation}\label{actionmimeticfraction}
S=\int \mathrm{d}x^4\left( \sqrt{-g}\left ( F(R)-V(\phi)+\eta \left(g^{\mu \nu}\partial_{\mu}\phi\partial_{\nu}\phi
+1\right)-\lambda \right )+\lambda \right )\, .
\end{equation}
By varying the action (\ref{actionmimeticfraction}) with respect to the metric $g_{\mu \nu}$, we obtain the following equations of motion, 
\begin{align}\label{eqnsofm1}
&\frac{g_{\mu \nu}}{2}\left(F(R)-V(\phi)+\eta \left(g^{\mu \nu}\partial_{\mu}\phi\partial_{\nu}\phi
+1\right)-\lambda  \right)-R_{\mu \nu}F'(R) -\eta \partial_{\mu}\phi\partial_{\nu}\phi+\nabla_{\mu}\nabla_{\nu}F'(R)-g_{\mu \nu}\square F'(R)=0\, ,
\end{align}
and also by varying the action (\ref{actionmimeticfraction}) with respect to the auxiliary field $\phi$, we obtain the following equation,
\begin{equation}\label{eqnsofm2}
-2\nabla^{\mu}\left( \eta \partial_{\mu}\phi\right)-V'(\phi)=0\, .
\end{equation}
It is conceivable that the variation of the $F(R)$ gravity action (\ref{actionmimeticfraction}), with respect to the Lagrange multiplier $\eta$, yields the mimetic constraint of Eq. (\ref{mimeticconstraint}), while variation with respect to $\lambda$, yields the unimodular constraint of Eq. (\ref{unimodularconstraint}). For the FRW background of Eq. (\ref{frw}), the $(t,t)$ components of (\ref{eqnsofm1}) is equal to,
\begin{equation}\label{enm1}
-F(R)+6(\dot{H}+H^2)F'(R)-6H\frac{\mathrm{d}F'(R)}{\mathrm{d}t}-\eta
(\dot{\phi}^2+1)+\lambda+V(\phi)=0\, ,
\end{equation}
while the $(i,j)$ components read,
\begin{equation}\label{enm2}
F(R)-2(\dot{H}+3H^2)+2\frac{\mathrm{d}^2F'(R)}{\mathrm{d}t^2}+4H\frac{\mathrm{d}F'(R)}{\mathrm{d}t}-\eta (\dot{\phi}^2-1)-V(\phi)-\lambda=0\, ,
\end{equation}
In addition, by using the metric (\ref{frw}) and the properties of the corresponding Christoffel symbols, Eq. (\ref{eqnsofm2}) takes the following form,
\begin{equation}\label{enm3}
2\frac{\mathrm{d}(\eta \dot{\phi})}{\mathrm{d}t}+6H\eta
\dot{\phi}-V'(\phi)=0\, .
\end{equation}
Finally, the mimetic constraint yields,
\begin{equation}\label{enm4}
\dot{\phi}^2-1=0\, .
\end{equation}
Note that in Eqs. (\ref{enm1}), (\ref{enm2}), (\ref{enm3}) and (\ref{enm4}), the ``prime'' denotes differentiation with respect to the Ricci scalar or with respect to the auxiliary scalar field $\phi$, while the ``dot'' denotes differentiation with respect to the cosmic time $t$. It easily follows from Eq. (\ref{enm4}), that the auxiliary scalar can be identified with the cosmic time, so $\phi=t$, and therefore, the equation of motion (\ref{enm1}) is simplified as follows,
\begin{equation}\label{enm1sim}
-F(R)+6(\dot{H}+H^2)F'(R)-6H\frac{\mathrm{d}F'(R)}{\mathrm{d}t}-2\eta+\lambda+V(\phi)=0\, ,
\end{equation}
while Eq. (\ref{enm2}) is simplified as follows,
\begin{equation}\label{enm2sim}
F(R)-2(\dot{H}+3H^2)+2\frac{\mathrm{d}^2F'(R)}{\mathrm{d}t^2}+4H\frac{\mathrm{d}F'(R)}{\mathrm{d}t}-V(\phi)-\lambda=0\,.
\end{equation}
Moreover, in view of the identification $\phi=t$, Eq. (\ref{enm4}) takes the following form,
\begin{equation}\label{onlali}
2\frac{\mathrm{d}\eta}{\mathrm{d}t}+6H\eta-V'(t)=0\, .
\end{equation}
By eliminating $\lambda$ from Eqs. (\ref{enm1sim}) and (\ref{enm2sim}), we obtain the following equation,
\begin{equation}\label{sone}
6(\dot{H}+H^2)F'(R)-2H\frac{\mathrm{d}F'(R)}{\mathrm{d}t}-2\eta-2(\dot{H}+3H^2)+2\frac{\mathrm{d}^2F'(R)}{\mathrm{d}t^2}-V(t)=0\, .
\end{equation}
By combining Eqs. (\ref{sone}) and (\ref{onlali}), we obtain the following first order differential equation,
\begin{equation}\label{auxeqn}
-2V'(t)-3H V(t)+f_0(t)=0\, ,
\end{equation}
where the function $f_0(t)$ stands for,
\begin{align}\label{explicitf0cosmictime}
& f_0(t)=-\Big{[} 18 H(t)\left( \dot{H}+H^2\right)F'(R)-6H^2\frac{\mathrm{d}F'(R)}{\mathrm{d}t}-6H\left(\dot{H}3H^2 \right)+6 H\frac{\mathrm{d}^3F'(R)}{\mathrm{d}t^3}\\ \notag &
6(\ddot{H}+2\dot{H}H)F'(R)+H\dot{H}\frac{\mathrm{d}F'(R)}{\mathrm{d}t}+6 H^2\frac{\mathrm{d}F'(R)}{\mathrm{d}t}
-2H\frac{\mathrm{d}^2F'(R)}{\mathrm{d}t^2}-2\left( \ddot{H}+6\dot{H}H\right)+2\frac{\mathrm{d}^2F'(R)}{\mathrm{d}t^2}\Big{]}\, .
\end{align}
The differential equation (\ref{auxeqn}) has the following general solution,
\begin{equation}\label{generalsol1}
V(t)=\frac{a^{3/2}(t)}{2}\int a^{-3/2}(t)f_0(t)\mathrm{d}t\, ,
\end{equation}
with $a(t)$ being the scale factor. Thus, by specifying the scale factor of an arbitrary cosmological evolution and the $F(R)$ gravity, by using Eq. (\ref{generalsol1}) we can obtain the mimetic potential that can generate such an evolution. Then, by using the resulting $V(t)$ potential and substituting it in Eq. (\ref{enm2sim}), we obtain the unimodular Lagrange multiplier function $\lambda$, responsible for the evolution $a(t)$. Finally, by substituting $\lambda$ and $V(t)$ in Eq. (\ref{enm1sim}), we obtain the mimetic Lagrange multiplier function $\eta$. Basically, we just described a quite general reconstruction method, which can realize quite arbitrary cosmological evolutions, given the scale factor and also the form of the $F(R)$ gravity. Also, given the form of the Lagrange multipliers $\lambda $ and $\eta$ and also the scale factor and the mimetic potential, we can solve the resulting differential equations to find which $F(R)$ gravity realizes such a cosmological evolution. In the rest of this section, by specifying $F(R)$ and $a(t)$, we investigate how the reconstruction method works in practise.

\subsection{Cosmological Scenarios from Unimodular Mimetic $F(R)$ Gravity}

In this section we use the reconstruction method we introduced in the previous section in order to realize two well known cosmological evolutions in the context of U-M $F(R)$ gravity. Our aim is to find the mimetic potential $V(t)$ and the Lagrange multiplier functions $\eta$ and $\lambda$, which can generate a given cosmological evolution with scale factor $a(t)$, for a specific given $F(R)$ gravity. The $F(R)$ gravity can have in principle an arbitrary form, so we shall consider the following very general form for the $F(R)$ gravity function,
\begin{equation}\label{generalfrmodels}
F(R)=R+\mu R^p+d R^q\, ,
\end{equation}
where the parameters $d,f,p,q$ are arbitrary real constant parameters. The class of $F(R)$ gravity models of the form (\ref{generalfrmodels}) is known to have quite interesting phenomenological implications, since it can generate late and early-time acceleration, see for example \cite{reviews1}, and also it can also result to very appealing astrophysical implications. In the rest of the paper, without loss of generality, we shall assume that the $F(R)$ gravity is of the form,
\begin{equation}\label{staro}
F(R)=R-d R^3+f R^2\, ,
\end{equation}
where the parameters $d$ and $f$ are assumed to be positive arbitrary real numbers. The $F(R)$ gravity model (\ref{staro}) leads to some interesting implications for neutron stars, since it generates an increase of the maximal neutron star mass, a result that cannot be obtained by the standard Einstein-Hilbert gravity. Having the interesting phenomenological implications of the model (\ref{staro}), we also demonstrate in this section that this model in the context of U-M gravity can also realize various cosmological evolutions.

\subsubsection{de Sitter Cosmology from U-M $F(R)$ Gravity}

We start off our analysis by studying the de Sitter realization in the context of U-M $F(R)$ gravity, so the scale factor and the Hubble rate are given below,
\begin{equation}\label{desitterscale}
a(t)=\e^{H_0\, t}\, ,\quad H(t)=H_0
\end{equation}
where $H_0$ is an arbitrary positive real parameter. Since we assumed that no matter fluids are present, by using Eqs. (\ref{generalsol1}) and (\ref{explicitf0cosmictime}), for the $F(R)$ gravity of Eq. (\ref{staro}), we easily find that the mimetic potential of the U-M $F(R)$ gravity for the de Sitter cosmology of Eq. (\ref{desitterscale}) is equal to $V(t)=0$. Therefore, by substituting in Eq. (\ref{enm2sim}), the unimodular Lagrange multiplier function $\lambda$ reads,
\begin{equation}\label{unilamdadesitter}
\lambda (t)=6 H_0^2 \left(1+24 f H_0^2-288 d H_0^4\right)\, ,
\end{equation}
so we can see that it is actually constant for the de Sitter cosmology case. Hence, by substituting the unimodular Lagrange multiplier $\lambda$ and the potential $V(t)$ in Eq. (\ref{enm1sim}), we can obtain the mimetic Lagrange multiplier $\eta$, which reads,
\begin{equation}\label{mimeticlagrangedesitter}
\eta (t)=6 H_0^2 \left(-1-12 f H_0^2+216 d H_0^4\right)\, ,
\end{equation}
which is also a constant parameter. So for the de Sitter cosmology case, the U-M $F(R)$ gravity Lagrangian has a very simple form. This result can also be compared with the corresponding ones in Refs. \cite{mimetic1} and \cite{unimime}, and it can be seen that the result in the case at hand is different.

\subsubsection{Perfect Fluid Cosmology from U-M Gravity}

As another example we consider the cosmology that is generated by a perfect fluid with a constant equation of state $p=w\rho$. The scale factor and the Hubble rate in this case are,
\begin{equation}\label{desitterscale1}
a(t)=t^{\frac{2}{3 (1+w)}}\, ,\quad H(t)=\frac{2}{3 t (1+w)}\, ,
\end{equation}
where $w$ stands for the equation of state parameter which is constant in this case. By substituting the scale factor and the Hubble rate (\ref{desitterscale1}) in Eq. (\ref{generalsol1}), and also for the $F(R)$ gravity (\ref{staro}), the mimetic potential can easily be found and the explicit form of which can be found in the Appendix, since it is lengthy to be presented here. Accordingly, the resulting forms of the Lagrange multipliers $\lambda$ and $\eta$ can also be found in the Appendix, and here we quote the exact form of the potential $V(t)$ and of $\lambda$ and $\eta$ for $w=1/3$, which corresponds to the radiation dominated evolution. Hence for $w=1/3$, the mimetic potential reads,
\begin{align}\label{mimeticpotdesitter1}
&V(t=\phi)= -\frac{2}{11 t^2}
\, ,
\end{align}
while the unimodular Lagrange multiplier function $\lambda (t)$, and the mimetic Lagrange multiplier $\eta$ read,
\begin{equation}\label{unilamdadesitter1}
\lambda (t)=-\frac{7}{22 t^2},\,\,\, \eta (t)=\frac{13}{22 t^2}\, .
\end{equation}
A direct comparison of the potential appearing in Eq. (\ref{mimeticpotdesitter1}) with the corresponding one of Ref. \cite{mukhanov2}, which is,
\begin{equation}\label{mimeperfectfluidcase}
V(t)=\frac{\mathcal{C}}{t^2}\, ,
\end{equation}
makes obvious that if the arbitrary constant in Eq. (\ref{mimeperfectfluidcase}) is chosen to be $\mathcal{C}=-2/11$, the potentials of Eqs. (\ref{mimeticpotdesitter1}) and (\ref{mimeperfectfluidcase}) become identical. Hence, we obtain the same cosmological evolution, for the same mimetic potential for the two theories, although in the case at hand, this result is obtained by using the $F(R)$ gravity (\ref{staro}).

\section{Unimodular Mimetic $F(R)$ Gravity Compatible with BICEP2/Keck Array Data}

Having described the general reconstruction method for the U-M $F(R)$ gravity, we now proceed to the realization of viable inflationary cosmologies from U-M $F(R)$ gravity. We assume that the $F(R)$ gravity has the functional form given in Eq. (\ref{staro}), and we use the perfect fluid approach \cite{perfectfluid} in order to calculate the slow-roll inflationary indices and also the corresponding observational indices. The perfect fluid formalism enables us to obtain the spectral indices in a model independent way and a detailed account on this formalism can be found in \cite{perfectfluid}. According to the perfect fluid approach, the $F(R)$ gravity is considered to be a perfect fluid, in which case, the slow-roll indices can be expressed as function of the Hubble rate as follows \cite{perfectfluid},
\begin{align}\label{hubbleslowrollnfolding}
&\epsilon=-\frac{H(N)}{4 H'(N)}\left(\frac{\frac{H''(N) }{H(N)}+6\frac{H'(N)
}{H(N)}+\left(\frac{H'(N)}{H(N)}\right)^2}{3+\frac{H'(N)}{H(N)}}\right)^2
\, ,\\ \notag &
\eta=-\frac{\left(9\frac{H'(N)}{H(N)}+3\frac{H''(N)}{H(N)}+\frac{1}{2}\left(
\frac{H'(N)}{H(N)}\right)^2-\frac{1}{2}\left(
\frac{H''(N)}{H'(N)}\right)^2+3
\frac{H''(N)}{H'(N)}+\frac{H'''(N)}{H'(N)}\right)}{2\left(3+\frac{H'(N)}{H(N
)}\right)}\, ,
\end{align}
where we used the $e$-foldings number $N$, instead of the cosmic time, and also the prime in Eq. (\ref{hubbleslowrollnfolding}) denotes differentiation with respect to the $e$-foldings number $N$. In addition, we used the following transformation properties,
\begin{equation}\label{transfefold}
\frac{\mathrm{d}}{\mathrm{d}t}=H(N)\frac{\mathrm{d}}{\mathrm{d}N},{\,}{\,}{\
,}
\frac{\mathrm{d}^2}{\mathrm{d}t^2}=H^2(N)\frac{\mathrm{d}^2}{\mathrm{d}N^2}+
H(N)\frac{\mathrm{d}H}{\mathrm{d}N}\frac{\mathrm{d}}{\mathrm{d}N}\, .
\end{equation}
Accordingly, the spectral index of primordial curvature perturbations $n_s$ and the scalar-to-tensor ratio $r$ in the perfect fluid approach, read,
\begin{equation}\label{indexspectrscratio}
n_s\simeq 1-6 \epsilon +2\eta,\, \, \, r=16\epsilon \, ,
\end{equation}
In order to perform a confrontation with the most recent observational data, and determine if the cosmological evolution can be considered as viable, here we quote the latest constraints on $n_s$ and $r$ coming from the Planck collaboration \cite{planck}, which indicate that,
\begin{equation}\label{constraintedvalues}
n_s=0.9644\pm 0.0049\, , \quad r<0.10\, ,
\end{equation}
and furthermore, the BICEP2/Keck Array data constrain the scalar to tensor ratio $r$ in the following way,
\begin{equation}\label{scalartotensorbicep2}
r<0.07\, ,
\end{equation}
A cosmological evolution that can be compatible with the observational data is the following,
\begin{equation}\label{hub1}
H(N)=\left(-G_0\text{  }N^{\beta }+G_1\right)^b\, ,
\end{equation}
where the parameters $G_0$, $G_1$, $\beta$ and $b$ are arbitrary real numbers. In order to avoid the issue that the Hubble rate turns negative, the parameter $b$ is assumed to have the following form,
\begin{equation}\label{bparameter}
b=\frac{2n}{2m+1},\,\,\, b<1\, ,
\end{equation}
where $m$ and $n$ are positive integers. Accordingly, by substituting Eq. (\ref{hub1}) in Eq. (\ref{hubbleslowrollnfolding}), the slow-roll parameter $\epsilon$ can easily be found and the same applies for the parameter $\eta$. Finally substituting their final form in the observational indices, the spectral index $n_s$ reads,
\begin{align}\label{observa}
& n_s=\frac{1}{2 N \Big{(}G_1-G_0 N^{\beta }\Big{)} \Big{(}-3 G_1 N+G_0
N^{\beta } (3 N+b \beta )\Big{)}^2}\times \\ \notag &
\Big{(}3 G_1^3 N \Big{(}-3+6 N (1+N)+4 \beta -6 N \beta -\beta
^2\Big{)}-G_0^3 N^{3 \beta } (9 N (-1+2 N (1+N)) \\ \notag &
 +48 b N^2 \beta +2 b^2 (-1+13 N) \beta ^2+4 b^3 \beta ^3\Big{)}-G_0 G_1^2
N^{\beta } \Big{(}54 N^3+2 b (-1+\beta ) \beta ^2+3 N (-1+\beta ) (9+\beta )
\\ \notag &  +6 N^2 (9-6 \beta +8 b \beta )\Big{)}+G_0^2 G_1 N^{2 \beta }
\Big{(}54 N^3+2 b (1+b) (-1+\beta ) \beta ^2 \\ \notag &
 +6 N^2 (9-3 \beta +16 b \beta )+N \Big{(}-27+2 \beta  \Big{(}6+3 \beta +13
b^2 \beta \Big{)}\Big{)}\Big{)}\Big{)} \, .
\end{align}
Assuming the following set of values for the free parameters of the theory $G_0$, $G_1$, $\beta$
and $b$,
\begin{equation}\label{defparam}
G_0=0.00009,\, \, \, G_1=400000, \, \, \,\beta=3.01, \, \, \, b=0.001\, ,
\end{equation}
and also assuming that $N=60$, we obtain the following values for $n_s$ and $r$,
\begin{equation}\label{scalindex}
n_s\simeq 0.966496, \, \, \, r\simeq 4.10971 \times 10^{-8}\, .
\end{equation}
A direct comparison of the values for $n_s$ and $r$ appearing in Eq. (\ref{scalindex}) with the Planck and BICEP2/Keck-Array data of Eqs. (\ref{constraintedvalues}) and (\ref{scalartotensorbicep2}), shows that both $n_s$ and $r$ are compatible with both the Planck and BICEP2/Keck-Array data. The cosmological evolution of Eq. (\ref{hub1}) can be realized in the context of U-M $F(R)$ gravity, by finding the mimetic potential $V(N)$ and correspondingly, the Lagrange multipliers $\eta (N)$ and $\lambda (N)$. Indeed, by using the transformation properties (\ref{transfefold}), the differential equation (\ref{auxeqn}) takes the following form,
\begin{equation}\label{difendnewnfolding}
-2H(N)\frac{\mathrm{d}V}{\mathrm{d}N}-3H(N)V(N)+f_0(N)=0\, ,
\end{equation}
where the function $f_0(N)$ is equal to,
\begin{align}\label{fnexplicitform}
& f_0(N)=-\Big{[}\left[18H\left(H\frac{\mathrm{d}H}{\mathrm{d}N}+H^2\right)+6\left(H^2\frac{\mathrm{d}^2H}{\mathrm{d}N^2}+H\left( \frac{\mathrm{d}}{\mathrm{d}N}\right)^2\right) \right]F'(R)\\ \notag &
H\frac{\mathrm{d}F'(R)}{\mathrm{d}N}\left( -6H^2+4H\frac{\mathrm{d}H}{\mathrm{d}N}6H^2\right)-2H\left(H^2\frac{\mathrm{d}^2F'(R)}{\mathrm{d}N^2} +H\frac{\mathrm{d}H}{\mathrm{d}N}\frac{\mathrm{d}F'(R)}{\mathrm{d}N}\right)+H\frac{\mathrm{d}\rho}{\mathrm{d}N}\\ \notag &
-2\left( H^2\frac{\mathrm{d}^2H}{\mathrm{d}N^2}+\left(\frac{\mathrm{d}H}{\mathrm{d}N}\right)^2\right)+2\left( H^2\frac{\mathrm{d}^2F'(R)}{\mathrm{d}N^2}+H\frac{\mathrm{d}H}{\mathrm{d}N}\frac{\mathrm{d}F'(R)}{\mathrm{d}N}\right)+\frac{\mathrm{d}P}{\mathrm{d}N}+3H\rho \\ \notag & -6 H\left( H \frac{\mathrm{d}H}{\mathrm{d}N}+3H^2\right)+3Hp+6H^2\frac{\mathrm{d}}{\mathrm{d}N}\left(H\frac{\mathrm{d}^2F'(R)}{\mathrm{d}N^2}+H \frac{\mathrm{d}H}{\mathrm{d}N}\frac{\mathrm{d}F'(R)}{\mathrm{d}N} \right)\Big{]}\, .
\end{align}
The general solution of the differential equation (\ref{difendnewnfolding}) is the following,
\begin{equation}\label{finalsoltuon}
V(N)=\frac{e^{-3N/2}}{2}\int e^{3N/2}f_0(N)\mathrm{d}N\, . 
\end{equation}
So by combining Eqs. (\ref{hub1}), (\ref{fnexplicitform}) and (\ref{finalsoltuon}) we may obtain the potential $V(N)$ and from it the Lagrange multipliers $\lambda (N)$ and $\eta (N)$ can be found, but we omit the details of this calculation for brevity.

\section{Unstable de Sitter Attractors and Graceful Exit from Inflation}

In this section we address the graceful exit issue that can arise for certain inflationary cosmologies in the context of U-M $F(R)$ gravity. The graceful exit can occur in a cosmological theory if unstable de Sitter vacua can be found in these theories. For an extended analysis on this issue see \cite{unimime} and the relevant references therein. In this paper we confine ourselves to finding the de Sitter vacua and proving that some of these are unstable. An unstable de Sitter attractor leads to the graceful exit from inflation, since the linear perturbations around these vacuum grow in an exponential way (in most cases). In section II we investigated how a de Sitter solution can be realized in the context of U-M $F(R)$ gravity, and as we evinced, the mimetic potential $V(\phi)$ was found to be equal to zero and in addition, the Lagrange multipliers $\eta (t)$ and $\lambda (t)$ are given in Eqs. (\ref{mimeticlagrangedesitter}) and (\ref{unilamdadesitter}) respectively. Our aim is to linearly perturb the de Sitter solution $H(t)=H_0$, so the solution will be of the form,
\begin{equation}\label{pertsol}
H(t)=H_0+\Delta H\, ,
\end{equation}
and insert this solution in the differential equation (\ref{enm1sim}). Then we keep only linear terms of $\Delta H$ and it's higher derivatives, by also using the fact that $V=0$ and that $\eta (t)$ and $\lambda (t)$ are given in Eqs. (\ref{mimeticlagrangedesitter}) and (\ref{unilamdadesitter}). But first, let us investigate how the solution $H=H_0$ can be expressed for the case that the $F(R)$ gravity is given by Eq. (\ref{staro}). By combining Eqs. (\ref{enm1sim}), (\ref{mimeticlagrangedesitter}) and (\ref{unilamdadesitter}), we obtain the following algebraic equation,
\begin{equation}\label{algebraiceqns}
12 H0^2 (1 + 24 f H0^2 - 432 d H0^4)=0\, ,
\end{equation}
which can easily be solved to yield the following de Sitter solutions,
\begin{equation}\label{skdes}
H_0=\frac{1}{6} \sqrt{\frac{f}{d}+\frac{\sqrt{3 d+f^2}}{d}},\,\,\,H_0=\frac{1}{6} \sqrt{\frac{f}{d}-\frac{\sqrt{3 d+f^2}}{d}}\, .
\end{equation}
Then by inserting Eq. (\ref{pertsol}) in the differential equation (\ref{enm1sim}), and by keeping linear terms of $\Delta H$ and it's higher order derivatives, we obtain the following differential equation obeyed by the perturbations,
\begin{align}\label{diffeqnnewpert}
& 12 H_0^2+288 f H_0^4-5184 d H_0^6-12 H_0 \Delta H(t)-5184 d H_0^5 \Delta H(t)\\ \notag & -216 d H_0^2 \Delta H'(t)+7776 d H_0^4 \Delta H'(t)-72 d H_0 \Delta H''(t)+2592 d H_0^3 \Delta H''(t)=0\, ,
\end{align}
where the prime indicates differentiation with respect to the cosmic time. The solution of this equation is,
\begin{equation}\label{soeqn1}
\Delta H(t)=\frac{-H_0-24 f H_0^3+432 d H_0^5}{-1-432 d H_0^4}+C_1e^{\mu_1t}+C_2e^{\mu_2 t}\, ,
\end{equation}
where $C_1$ and $C_2$ are arbitrary integration parameters, and also the parameters $\mu_1$ and $\mu_2$ are equal to,
\begin{align}\label{mu1mu2}
& \mu_1=\frac{\left(9 d H_0-324 d H_0^3+\sqrt{3} \sqrt{-2 d+72 d H_0^2+27 d^2 H_0^2-2808 d^2 H_0^4+66096 d^2 H_0^6}\right)}{6 \left(-d+36 d H_0^2\right)},\,\,\,\\ \notag & \mu_2=\frac{\left(9 d H_0-324 d H_0^3-\sqrt{3} \sqrt{-2 d+72 d H_0^2+27 d^2 H_0^2-2808 d^2 H_0^4+66096 d^2 H_0^6}\right) t}{6 \left(-d+36 d H_0^2\right)}\,.
\end{align}
By substituting the second solution of Eq. (\ref{skdes}) in $\mu_2$, it can be seen that $\mu_2>0$ for all the values of $d$ and $f$, and hence the perturbation $\Delta H(t)$ grows as a function of time in an exponential way\footnote{for example, for $d=10^{-6}$ and $f=100$, the parameter $\mu_2$ reads $\mu_2=1.3\times 10^6$.}. Therefore, the de Sitter solution is unstable and hence graceful exit can be achieved by the growing curvature perturbations. For more details on this mechanism for graceful exit and its connection with the final attractor theorem, see Ref. \cite{unimime}.

\section{Conclusions}

In this paper we presented the extension of unimodular-mimetic gravity which was developed in Ref. \cite{unimime} in the context of $F(R)$ gravity. After developing the formalism of unimodular-mimetic $F(R)$ gravity, we demonstrated that the resulting equations of motion constitute a reconstruction method with which it is possible to realize various cosmological scenarios. Particularly, the reconstruction method can work in various ways, for example, if the cosmological evolution and the $F(R)$ gravity are given, then it is possible to find which unimodular and mimetic Lagrange multipliers generate such a cosmological evolution. In addition, if the cosmological evolution and the potential along with the Lagrange multipliers are fixed, then it is possible to find which $F(R)$ gravity generates the specific cosmology under study. We used the first approach and we investigated which unimodular $F(R)$ gravity can realize the de Sitter and perfect fluid cosmology. Particularly, since the choice of the $F(R)$ gravity can be arbitrarily done, we chose an $F(R)$ gravity that has some phenomenological significance, with regards to the astrophysical implications that it generates, and for this $F(R)$ gravity, we investigated which unimodular and mimetic Lagrange multipliers and also which mimetic potential, can generate the specific cosmology at hand. The cosmological evolutions we studied can be realized by the standard Einstein-Hilbert gravity, but the advantage of the unimodular $F(R)$ gravity approach is that even exotic cosmological scenarios for the standard Einstein-Hilbert gravity, can be consistently realized from unimodular-mimetic $F(R)$ gravity. Also, we demonstrated how some inflationary cosmologies can be described by unimodular-mimetic $F(R)$ gravity, and we found that compatibility with the latest observational data can be achieved. We also addressed the graceful exit issue and we showed that in the unimodular-mimetic $F(R)$ gravity, graceful exit can be achieved via unstable de Sitter vacua, since these lead to growing curvature perturbations. The theoretical framework of unimodular-mimetic $F(R)$ gravity we presented in this paper, can easily be extended to other modified gravity scenarios, such as $F(R,G)$ gravity, where $G$ stands for the Gauss-Bonnet invariant, so we leave this for a future work.

\section*{Acknowledgments}

This work is supported in part by MINECO (Spain), project FIS2013-44881
(S.D.O) and partly by Min. of Education and Science of Russia (S.D.O and
V.K.O).

\section*{Appendix:Explicit Forms of the Mimetic Potential and Lagrange Multipliers}

Here we present the explicit forms of the mimetic potential and of the Lagrange multipliers for various cases appearing in the main text of the paper. We start off with the cosmological evolution (\ref{desitterscale1}), in which case the mimetic potential reads,
\begin{align}\label{mimeticpotdesitter1app}
&V(t=\phi)= \left(27 t^7 (1+w)^5 (3+2 w) (4+3 w) (5+4 w) (6+5 w) (7+6 w) (8+7 w)\right)^{-1}\times \\ \notag &
-8 \left(-8 d (1-3 w)^2 (3+2 w) (4+3 w) (5+4 w) \left(45 t^2 (1+w)^2 (7+6 w) (8+7 w)-3 t (6+5 w) (8+7 w) (9+7 w)\right.\right.\\ \notag &
-2 (6+5 w) (7+6 w) (269+270 w))+t^2 (1+w)^2 (7+6 w) (8+7 w)\\ \notag &
\left(9 t^3 w (1+w)^2 (4+3 w) (5+4 w) (6+5 w)-2 f (-3+w (7+6 w)) \right. \\ \notag &
\left.\left.\left(-12 t (4+3 w) (6+5 w)+27 t^2 (1+w)^2 (5+4 w) (6+5 w)-2 (4+3 w) (5+4 w) (107+108 w)\right)\right)\right)
\, ,
\end{align}
and in addition, by using Eq. (\ref{mimeticlagrangedesitter1app}), the corresponding unimodular Lagrange multiplier function $\lambda (t)$ reads,
\begin{align}\label{unilamdadesitter1app}
& \lambda (t)=\left(27 t^7 (1+w)^6 (3+2 w) (4+3 w) (5+4 w) (6+5 w) (7+6 w) (8+7 w)\right)\times \\ \notag &
-\left(8 \left(8 d (1-3 w)^2 \left(60+133 w+98 w^2+24 w^3\right) \left(45 t^2 (1+w)^3 -2 \left(11298+41737 w+57679 w^2+35340 w^3+8100 w^4\right)\right.\right.\right.\\ \notag &
\left.+t \left(21216+104644 w+205616 w^2+201009 w^3+97725 w^4+18900 w^5\right)\right)t^2 (1+w)^2 \left(56+97 w+42 w^2\right)\\ \notag &
\left(9 t^3 w (1+w)^2 \left(240+692 w+740 w^2+347 w^3+60 w^4\right)f \left(-3+7 w+6 w^2\right)\right.\\ \notag &
\left(27 t^2 (1+w)^3 \left(30+49 w+20 w^2\right)-2 \left(2140+7617 w+10109 w^2+5928 w^3+1296 w^4\right)\right.\\ \notag &
\left.\left.\left.\left.+3 t \left(1224+5898 w+11595 w^2+11423 w^3+5586 w^4+1080 w^5\right)\right)\right)\right)\right)
\end{align}
Correspondingly, by combining Eqs.~(\ref{mimeticpotdesitter1app}) and (\ref{unilamdadesitter1app}), the mimetic Lagrange multiplier $\eta$ reads,
\begin{align}\label{mimeticlagrangedesitter1app}
& \eta (t)=\left(27 t^7 (1+w)^6 (3+2 w) (4+3 w) (5+4 w) (6+5 w) (7+6 w) (8+7 w)\right)^{-1}\times \\ \notag &
-\left(4 \left(8 d (1-3 w)^2 (3+2 w) (4+3 w) (5+4 w) \left(45 t^2 (1+w)^3 (7+6 w) (8+7 w)-2 (1+w) (6+5 w) (7+6 w) (269+270 w)\right.\right.\right.\\ \notag &
+3 t (6+5 w) (8+7 w) (236+w (663+5 w (121+36 w))))t^2 (1+w)^2 (7+6 w) (8+7 w)\\ \notag &
\left(-9 t^3 w (1+w)^2 (4+3 w) (5+4 w) (6+5 w) (7+5 w)+2 f (-3+w (7+6 w)) \left(27 t^2 (1+w)^3 (5+4 w) (6+5 w)\right.\right.\\ \notag &
+6 t (4+3 w) (6+5 w) (58+w (161+w (137+36 w)))))))
\end{align}


\end{document}